# Magnetic Switching of Phase-Slip Dissipation in NbSe$_2$ Nanobelts

Abram Falk[1†], Mandar M. Deshmukh[1†], Amy L. Prieto[2], Jeffrey J. Urban[2], Andrea Jonas[2], Hongkun Park[1,2*]

*[1]Department of Physics and [2]Department of Chemistry and Chemical Biology, Harvard University, 12 Oxford St., Cambridge, MA 02138, USA*

[†]These authors contributed equally to this work.

[*]To whom correspondence should be addressed. E-mail: Hongkun_Park@harvard.edu

The stability of the superconducting dissipationless and resistive states in single-crystalline NbSe$_2$ nanobelts is characterized by transport measurements in an external magnetic field (H). Current-driven electrical measurements show voltage steps, indicating the nucleation of phase-slip structures. Well below the critical temperature, the position of the voltage steps exhibits a sharp, periodic dependence as a function of H. This phenomenon is discussed in the context of two possible mechanisms: the interference of the order parameter and the periodic rearrangement of the vortex lattice within the nanobelt.

One-dimensional (1D) superconductors refer to wire-like superconducting materials whose thickness and width are smaller than the Ginsburg-Landau coherence length ($\xi$) and magnetic penetration depth *($\lambda$)*. These materials have a uniform current distribution over their cross-sections,[1] and the onset of resistance in the current (*I*)-biasing condition is characterized by a sequence of regular voltage (*V*) steps, corresponding to the nucleation of phase-slip centers (PSCs).[2, 3] Recent advances in the synthesis of nanostructured materials have yielded a variety of 1D superconductors, including electrodeposited Pb and Sn nanowires,[4] nanotube-templated amorphous $MoGe$[5] and $Nb$[6] nanowires, lithographically-defined[7] and laser-ablated[8] $YBa_2Cu_3O_{7-\delta}$ nanowires, and $MgB_2$ nanowires.[9]

Mesoscopic wires and strips, with one or more dimensions wider than $\xi$, can also exhibit PSC-like *V* steps, attributable to Josephson weak-links and spatially-extended PSCs.[10] Mesoscopic superconducting wires are also thick enough to support vortices, and thus they provide an interesting model system for investigating the coexistence and the interplay between the PSC-like structures (PSSs) and the Abrikosov vortex lattice.[11] Although mesoscopic $NbSe_2$ films have been used to study single-vortex physics in absence of an applied *I*,[12] *I*-driven few-vortex physics has never been investigated in superconducting nanowires.

Here we report transport studies of individual $NbSe_2$ nanobelts that explore the effect of an external magnetic field on the stability of PSSs. The *I-V* characteristic of the nanobelts displays a series of *V* steps attributable to nucleating PSSs. At low temperatures, the *I-V* curves become hysteretic. When an external magnetic field (**H**) is applied, the hysteresis sharply decreases for small windows at periodic values of **H**, indicating the switching on and off of phase-slip dissipation caused by thermal activation. We discuss this phenomenon in the context of two possible mechanisms, the interference of the order parameter and the periodic rearrangement of the vortex lattice within the nanobelt.

Single-crystalline $NbSe_2$ nanobelts were synthesized by annealing $NbSe_3$ nanobelts in an inert gas and converting them into $NbSe_2$ nanobelts.[13] Transmission electron microscopy and X-ray diffraction studies show that the $NbSe_2$ nanobelts are single crystalline and have a ribbon-like geometry[14] with the *c*-axis perpendicular to the axis of the nanobelt (Fig. 1). The thickness (*d*) and width (*w*) of typical $NbSe_2$ nanobelts are ~100 nm, indicating that $\xi(T = 0) < d, w \sim \lambda \, (T = 0)$.[15, 16]

Devices incorporating individual $NbSe_2$ nanobelts were fabricated by suspending the nanobelts in isopropanol, depositing them on a $SiO_x$ or AlN substrate, and contacting them to bonding pads using conventional electron-beam lithography followed by sputtering of 60-nm Cr and 200-nm Au layers. Immediately before sputtering, the nanobelts were cleaned with an Ar plasma in the sputtering chamber. This procedure allows the formation of metal-nanobelt junctions with low contact resistance (20 ~ 100 $\Omega$).

Transport measurements were performed using either a vapor-cooled variable-temperature insert or a 300mK helium-3 insert in an Oxford Instruments cryostat. The resistance of $NbSe_2$ nanobelts were measured by *I*-biasing the nanobelts with a home-built current source and then measuring the four-probe *V*. Differential resistance (*dV/dI*) of the device was also recorded simultaneously using a lock-in amplifier set at frequency *f* = 1 kHz. Measurements on multiple $NbSe_2$ nanobelt samples showed that the superconducting transition temperature $T_C$ of NbSe$_2$ nanobelts typically ranged between 2 to 2.3 K, lower than bulk values ($T_{C,bulk}$ = 7 K).[16] Previous studies have shown that Se impurities resulting from the incomplete conversion of NbSe$_3$ to NbSe$_2$ suppresses $T_C$.[16]

Figure 2(a) shows the basic *I-V* characteristic of device A at $T$ = 1.7 K ($T_C$ = 2.1 K for this nanobelt). At low $I$, the $V$ drop across the nanobelt is small but nonzero and increases with $I$. As $I$ increases above a switching current ($I_{sw}$), the first $V$ step appears followed by several more. The *I-V* curves are reproducible and, at this temperature, non-hysteretic. The linear regions between the $V$ steps roughly extrapolate to a single point, $I_s$ = 4.2 ± 0.5 µA. At high $I$ ($I$ > 40 µA), the *I-V* characteristic curves toward the asymptote $V = R_N I$, where $R_N$ is the normal-state resistance of the nanobelt, indicating that the nanobelt becomes normal.

The steps in $V$ appear as peaks in *dV/dI*. Figure 2(b) shows the evolution of these peaks as a function of $T$. As $T$ increases, the positions of the $V$ steps gradually move inward toward zero current, their width and magnitude soften, and a single *I-V* step occasionally branches into two distinct steps. These features are symmetric across $I = 0$, again confirming the absence of *I-V* hysteresis at $T \geq$ 1.7 K.

The inset to Fig. 2(a) shows the *I-V* characteristic of device A at $T$ = 300 mK, and shows that hysteresis emerges at lower temperatures. Specifically, the positions of the innermost $V$ step for the $I$ sweep-up ($I_{sw}$) and sweep-down directions (retrapping current $I_r$) differ significantly from each other ($I_{sw} > I_r$). Additional $V$ steps also appear in the range $I_{sw} < I < I_r$ during the $I$ down-sweep. The device behavior before the first $V$ step is characterized by a dissipationless regime ($0 < I < I_c$) and a low-resistance regime ($I_c < I <$ $I_{sw}$). The low-resistance regime is too small to be visible in the Fig. 2(a) inset, but it shows up clearly in Fig. 3.

The investigation of the device *I-V* characteristics at nonzero **H** reveals an intriguing feature that is periodic in **H**. Specifically, Fig. 3 shows the evolution of d$V$/d$I$ from device B as a function of **H**, where **H** ∥ **b** (see Fig. 1 for the definition of the crystalline axes). As H is increased, $I_r$(H) is periodically enhanced with period ΔH = 270 ± 30 mT. These enhancements have a finite width, δH = 35 ± 5 mT and a binary on-off character, at least within the resolution of our scan (5 mT). The boost to $I_r$ at these values of **H** is accompanied by a diminishment to $I_{sw}$, although the effect of **H** on $I_{sw}$ is generally smaller and more varied than its effect on $I_r$. The other steps do not display the periodic variations, although they do gradually move inward as an increasing **H** suppresses the order parameter.

The periodic features illustrated in Fig. 3 were observed in 5 out of the 7 devices that have been measured in detail. Moreover, Fig. 4 shows that these periodic features appear regardless of the **H** orientation with respect to the nanobelt axis, although not every device showed the periodic feature for every orientation of **H**. Comparison of the field scale in Figs. 4(a) and 4(b) indicates that the misalignment of **H** with respect to the nanobelt axis cannot account for the H-periodic features for three different **H** orientations. As Fig. 4(b) shows, for a given orientation of **H**, the evolution of the *I-V* curves is symmetric about H = 0.

The *V* steps in Figs. 2 and 3 are consistent with the nucleation of PSSs observed in low-dimensional superconductors.[3] Specifically, the observation that the linear regions between the voltage steps extrapolate to the point $I_s$ indicates that the *V* steps observed in NbSe$_2$ nanobelt devices behave according to the PSC model proposed by Skocpol, Beasley, and Tinkham (SBT).[17] According to the SBT model, the height of the PSC $V$ step is given by

$$V_{step} = \frac{2\Lambda R_N (I - I_s)}{L},$$

where $\Lambda$ is the quasiparticle diffusion length that signifies the length scale over which the electrostatic potential and the supercurrent density vary. The fit of the data in Fig. 2 to the SBT model yields $\Lambda \sim 4$ μm. Since our NbSe$_2$ nanobelts are too thick to be truly 1D superconductors, the observed PSSs may have internal structures, such as channels of flowing Abrikosov or kinetic vortices,[18] that arise from the variation of the order parameter over the cross-section of the superconductor.

The $I$-$V$ hysteresis observed in Fig. 3 follows from the theory of underdamped resistively and capacitively shunted Josephson junctions (RCSJ) in the presence of thermal noise.[19] A PSS is a temporal oscillation of the order parameter and thus can be considered as a dynamically generated Josephson junction.[1] The importance of thermal fluctuations in Josephson junctions is measured by the parameter $\Gamma = k_B T / E_J$, where the Josephson energy $E_J$ is given by $E_J = \hbar I_{sw0}/2e$, and $I_{sw0}$ is the switching current at $T = 0$. In the RCSJ model, thermal fluctuations can excite the junction out of the dissipationless state and into its resistive state, effectively reducing $I_{sw}$. When an under-damped junction is in its resistive state and $I$ is being ramped down, on the other hand, thermal fluctuations can prematurely trap the junction into its dissipationless state, increasing $I_r$. The periodic feature in our data, which appears as a boost to $I_r$ accompanied by a decrease in $I_{sw}$, is thus explained by a periodic increase in $\Gamma$ with H.

Within the RCSJ model, an increase in $\Gamma$ in our $NbSe_2$ nanobelt devices should derive from a decrease in $E_J$, because $T$ is held constant as H is varied.[20] Unfortunately, the experimental data in Figs. 2 and 3 by themselves do not provide a microscopic insight into the underlying cause of the periodic variation in $E_J$. One possible mechanism is the interference of the order parameter caused by the magnetic field threading through the PSS, leading to an oscillatory $E_J(H)$ behavior.[21] The matching of vortex rows to the Bean-Livingston surface barrier[22] can also explain such a periodic $E_J$ dip. In this latter scenario, vortices fill up the thin film row-by-row as H is increased from zero. At certain special values of H called matching fields, a complete chain of in-plane vortices will be filled, and the vortex lattice rearranges to form a new row. If we identify the current at which the vortex lattice becomes unstable with $I_{sw0}$, then the instability of the vortex lattice corresponds to a minimum in $E_J$. Unfortunately, neither of these mechanisms accounts for the "on-off" nature of the periodic feature in Fig. 3. Figure 4(c) shows that the periodically enhanced $I_r$ coincides with the line delineated by the second innermost PSC, suggesting that PSC synchronization[6, 23] might play a role.

Because the microscopic mechanism for the $E_J$ reduction remains poorly understood, so does the physical meaning of $\delta H$ and $\Delta H$. When **H** is parallel to the nanobelt axis, however, it is natural to normalize these quantities to $w$, $d$, and the flux quantum $\phi_0$, effectively counting the number of vortices that penetrate the nanobelt. For device B (C) shown in Fig 4(b) (Fig 4(d)), the period of the $I_r$ enhancement is $\Delta H = 500 \pm 20$ mT (700 $\pm 200$ mT), corresponding to the addition of ~11 vortices, and the width of a single feature is $\delta H = 50 \pm 10$ mT ($60 \pm 20$ mT), corresponding to the addition of $1.0 \pm 0.3$ vortices. The fact that the normalized $\delta H$ approximately corresponds to one vortex suggests that single or few vortex processes may play a role in the magnetic switching phenomenon.

The transport study of $NbSe_2$ nanobelts presented here reveals that the values of $I_{sw}$ and $I_r$, which measure the stability of the superconducting dissipationless and resistive states, respectively, switch sharply and periodically as functions of H. This observation indicates the periodic variation of $E_J$ with H. Our results suggest that superconducting nanobelts can be an interesting model system for illuminating the interplay between phase-slip dissipation and $I$-driven vortex physics.

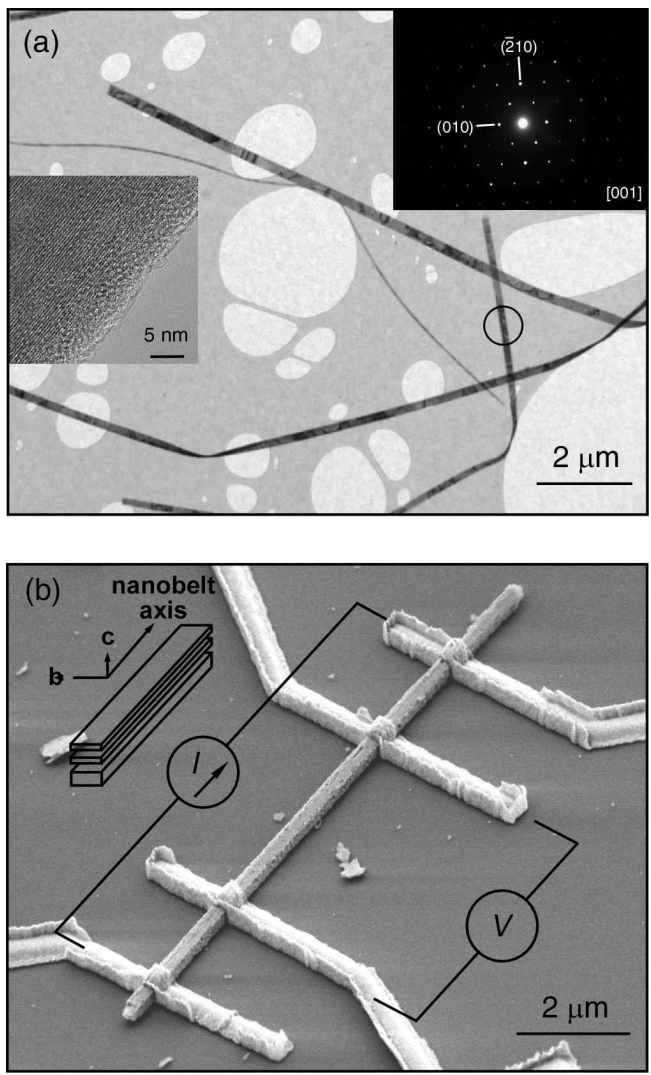

**FIG. 1.** (a) Transmission electron micrograph of NbSe$_2$ nanobelts. Inset (left): a high-resolution transmission electron micrograph of a nanobelt. Inset (upper-right): an electron diffraction pattern of the circled nanobelt. (b) A scanning electron micrograph of a typical four-probe device.

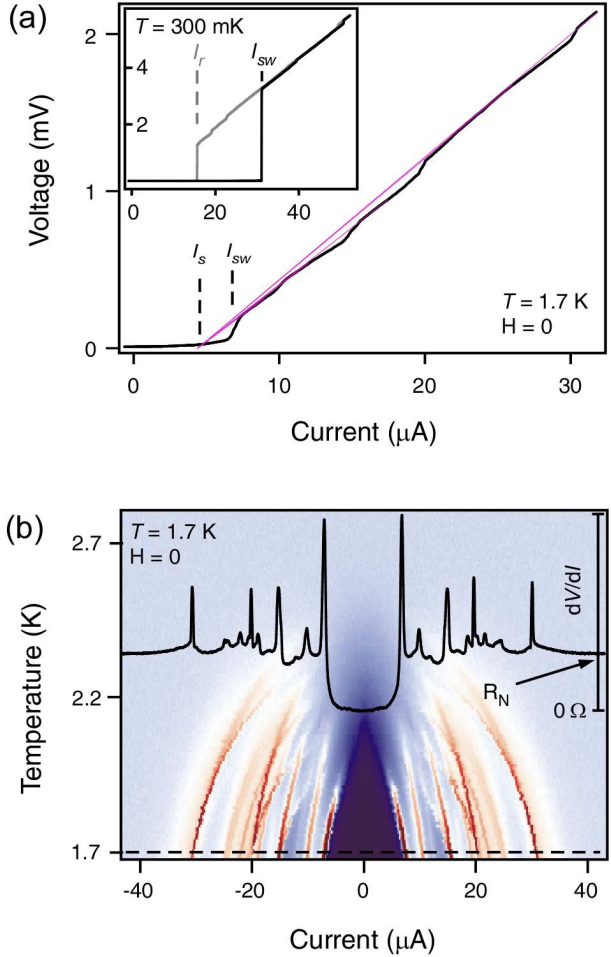

**FIG. 2.** (a) The *I-V* curve of device A at *T* = 1.7 K and **H** = 0. The pink lines are a constrained fit, illustrating how the linear portions of the *I-V* curve between voltage steps extrapolate to the point $I_s$. Inset (upper left): The *I-V* curve of device A at *T* = 300 mK and **H** = 0. The dimensions (*w × d × inner electrode spacing*) of device A are 130 × 430 nm × 10.1 µm. (b) d*V*/d*I* of device A as a function of *T* and *I*. The color scale changes from purple (0), black, dark blue, light blue ($R_N$), white, to red. The current sweep direction is left-to-right. The line graph is a cross-section of d*V*/d*I* along the dashed line at *T* = 1.7 K.

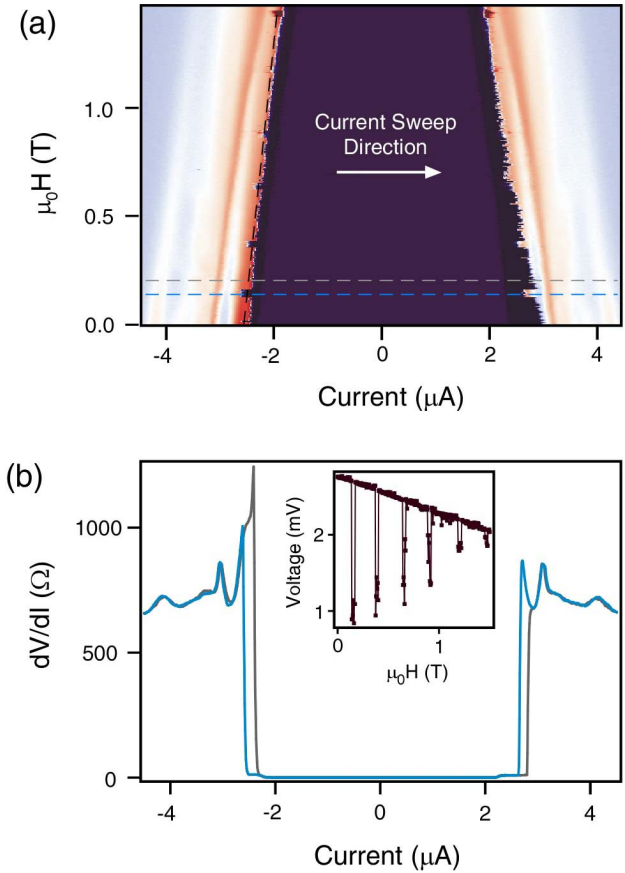

**FIG 3.** (a) d$V$/d$I$ of device B plotted as a function of H ( **H** ‖ **b** ) and $I$, at $T$ = 300 mK. (b) Two horizontal cross-sections of Fig. 3(a), illustrating the decrease of hysteresis in certain periodic windows of H. Inset: A diagonal cross-section of Fig. 3(a) along the dashed line, illustrating the periodic magnetic switching effect. Device B has dimensions 90 × 530 nm × 7.3 μm.

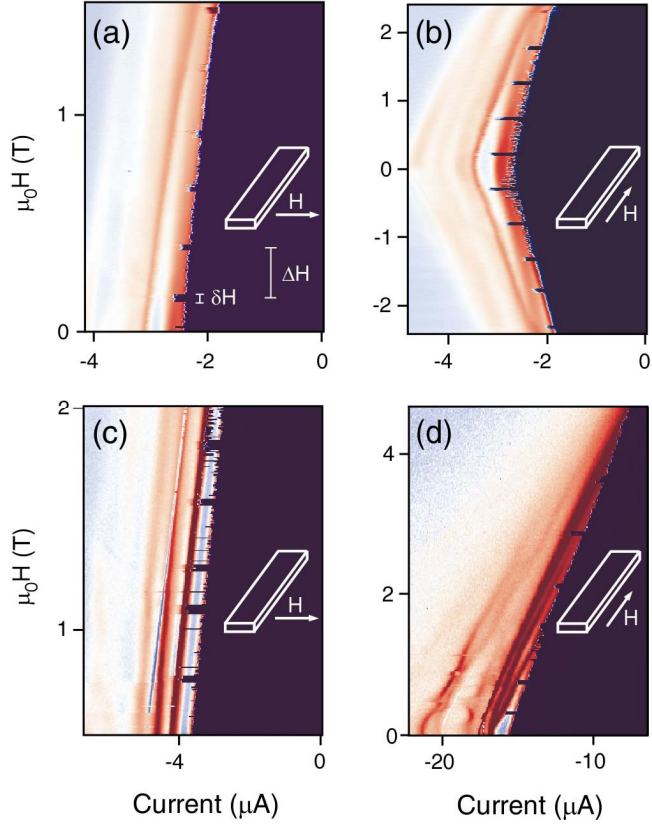

**FIG. 4.** d$V$/d$I$ plotted as a function of **H** and $I$, at $T$ = 300 mK, of (a) device B with

**H** ∥ **b** , (b) device B with **H** ∥ nanowire axis, (c) device C with **H** ∥ **b** , and (d) device D

with **H** ∥ nanowire axis. The negative $I$ values indicate that $I$ is being swept down to zero

(left-to-right). The dimensions of device C are 60 × 525 nm × 3.9 μm, and device D is 65

× 375 nm × 4.8 μm.